\documentclass[preprint,11pt]{elsarticle}
\usepackage[T1]{fontenc}% optional T1 font encoding
\usepackage[table,xcdraw]{xcolor}
\usepackage{float}
\usepackage{tabularx,color}
\usepackage{listings}
\usepackage{multirow}
\usepackage{lipsum}
\usepackage[nolist,nohyperlinks]{acronym}
\usepackage[hidelinks, bookmarks=false, draft]{hyperref}
\usepackage{graphicx} 
\usepackage{pdfpages}			
\usepackage{times}              
\usepackage[ruled,vlined]{algorithm2e}
\usepackage{lipsum} 
\usepackage{epstopdf}
\usepackage{array}
\usepackage{subfigure}
\usepackage{pbox}
\usepackage{blindtext}
\usepackage{booktabs}
\usepackage{subfigure}
\usepackage[utf8]{inputenc}
\usepackage{verbatim}
\usepackage{ragged2e}

\usepackage{hyphenat}
\usepackage{amssymb}
\usepackage{xspace}
\usepackage{lineno}
\usepackage{color}
\usepackage[hidelinks, bookmarks=false, draft]{hyperref}

\journal{Future Generation Computer Systems}

\begin{document}

\begin{frontmatter}

\title{
%System Intelligence to Mitigate Connectivity Issues in UAV-Based Communications\\
% UAV-Based Critical Missions with System Intelligence for Challenging Connectivity
System Intelligence for UAV-Based Mission Critical with Challenging 5G/B5G Connectivity}

\author[mymainaddress]{Cristiano Bonato Both}
\address[mymainaddress]{University of Vale do Rio dos Sinos - UNISINOS - São Leopoldo, Brazil}
\cortext[mycorrespondingauthor]{Corresponding author}
\ead{cbboth@unisinos.br}

\author[secondaddress]{João Borges}
\author[secondaddress]{Luan Gonçalves}
\author[secondaddress]{Cleverson Nahum}
\address[secondaddress]{Federal University of Pará - UFPA - Belém, Brazil}

\author[thirdaddress]{Ciro Macedo}
\address[thirdaddress]{Federal University of Goiás - UFG - Goiânia, Brazil}

\author[secondaddress]{Aldebaro Klautau}

\author[thirdaddress]{Kleber Cardoso}

\begin{abstract}
%\lipsum[1]
%UAV swarm in 5G and beyond
%New architectures and communication protocols for cellular-connected UAVs
%the System Intelligence (SI) as 
Unmanned aerial vehicles (UAVs) and communication systems are fundamental elements in Mission Critical services, such as search and rescue. In this article, we introduce an architecture for managing and orchestrating 5G and beyond networks that operate over a heterogeneous infrastructure with UAVs' aid. UAVs are used for collecting and processing data, as well as improving communications. The proposed System Intelligence (SI) architecture was designed to comply with recent standardization works, especially the ETSI Experiential Networked Intelligence specifications. 
%This article's contribution
Another contribution of this article is an evaluation using a testbed based on a virtualized non-standalone 5G core and a 4G Radio Access Network (RAN) implemented with open-source software. The experimental results indicate, for instance, that SI can substantially improve the latency of UAV-based services by splitting deep neural networks between UAV and edge or cloud equipment. Other experiments explore the slicing of RAN resources and efficient placement of virtual network functions to assess the benefits of incorporating intelligence in UAV-based mission critical services.
\end{abstract}


\end{frontmatter}

%\author{Cristiano Both,
%João Borges,
%Luan Gonçalves,
%Cleverson Nahum, \\
%Ciro Macedo,
%Aldebaro Klautau, and
%Kleber Cardoso
%\thanks{Cristiano Both is with the University of Vale do Rio dos Sinos; 
%Aldebaro Klautau, João Borges, Luan Gonçalves, and Cleverson Nahum are with the Federal University of Pará; 
%Ciro Macedo and Kleber Cardoso are with Universidade Federal de Goiás.
%}% <-this % stops a space
%}

%\clearpage\maketitle
%\thispagestyle{empty}

\section{Introduction}

%\ak{Mission Critical (MC) or Critical Missions as in the title?}

%\kc{I suggest Mission Critical (MC), \ac{SI}nce this is the term adopted by 3GPP: \url{https://www.3gpp.org/news-events/1875-mc} services and also by the ITU: \url{https://www.itu.int/en/myitu/News/2020/05/11/13/21/Mission-critical-visual-situational-awareness-in-times-of-disasters}}

Annually, the economic losses from global disasters sum hundreds of billions of dollars\footnote{{https://ourworldindata.org/grapher/economic-damage-from-natural-disasters}} and reap thousands of human lives. The impact on wildlife has more diffuse numbers~\cite{iwasaki2018}, but all of them expose a dramatic situation. Communication systems are fundamental tools in these scenarios, since they provide the means for proper coordination among the several teams involved in diverse tasks, from saving lives to repairing infrastructures. More recently, \acp{UAV} have been added as another valuable tools in the context of Mission Critical (MC) services, offering help on localizing people and animals, transporting medicines, improving communications, among other relevant tasks~\cite{AutoSOS}. For example, \ac{SAR} missions are services usually performed in remote areas where the telecommunication infrastructure is inoperative or severely damaged due to disasters, and UAVs can be extremely helpful. However, the deployment and operation of \acp{UAV} in the \ac{MC} context still depends strongly in human intervention, what limits its applicability and efficiency. Similar problem is observed with communication systems, mainly involving data transport, which is becoming the most relevant even in disasters scenarios. Considering these problems, standardization and adoption of \ac{AI/ML} solutions are promising approach to tackle these issues \cite{Bonati-2020}.

In general, the investments in \ac{MC} scenarios' resources 
%have an object to 
aim at preventing losses, not in promoting %earnings. 
profit.
Therefore, it is essential to increase scale and reduce costs in any \ac{MC}-related solution. The adoption of worldwide standards is a promising approach to achieve this goal. Additionally, international cooperation in \ac{MC} scenarios is quite common and standards become again a natural choice. Nowadays, the \ac{5G} and its evolution, i.e., \ac{B5G}, are a key set of standards in the general context and also in the \ac{MC} scenarios. \ac{MC} communications have been a concern in the standardization bodies since long time, however, the focus has been in a specific set of \ac{MC} services considered critical to teams involved in \ac{MC} scenario. Additionally, the support for \ac{UAV} is very recent, only considered effectively in the last \ac{3GPP} Releases, 15 \cite{3gpp:rel15nr21.915} and 16 \cite{3gpp:rel16nr21.916}. Nevertheless, these standards assume \acp{UAV} acting mainly as \ac{UE} and commanded by human beings. While fleets of autonomous \acp{UAV}, able to deploy and operate fully functional networks and services, have been investigated in the literature~\cite{mozaffari:19, Pham-2020, Cazzato-2020}, this context is still considered futuristic and out of scope for the standardization bodies in the near future. In this article, we argue that \ac{AI/ML} can turn this vision into reality sooner than is being expected.

\ac{AI/ML} is already being widely adopted in \ac{UAV} for autonomous operation \cite{Shafin-2020}. Moreover, the standardization bodies of communication systems are in an advanced stage of defining how to adopt and deploy \ac{AI/ML} \cite{ENI-2020}. However, the main efforts are focused on traditional infrastructure, i.e., without considering \ac{UAV} as part of the whole system and acting not only as \ac{UE} but also as part of the Radio Access Network (RAN) and even the \ac{5GC}. Furthermore, \acp{UAV} can offer edge computing resources, which enables several new applications and services. Finally, \acp{UAV} can provide temporary communication systems that are fully functional end-to-end. Software systems such as \ac{5GC} and the virtualized and disaggregated \ac{RAN} are receiving \ac{AI/ML} updates to decrease the need for human intervention. While these updates may be enough for traditional infrastructure, they cannot deal with the challenging conditions faced by \acp{UAV} in MC scenarios. Additionally, each \ac{AI/ML} solution is limited to its related system, disregarding the broader context that involves multiple interconnected systems, e.g., \ac{5GC}, \ac{RAN}, and \ac{MEC}.

The contributions of this article are two-fold: (i) we introduce the \ac{SI} as an architecture for managing and orchestrating 5G/B5G communication systems that operate over a heterogeneous infrastructure, which includes \acp{UAV}. Besides describing \ac{SI} and contextualizing it concerning existing standards, (ii) this article also contributes with results obtained with a testbed that supports experiments with \ac{UAV}-based communications and  AI in future networks and computer systems.

The proposed \ac{SI} must support advanced MC services in several scenarios, including those in which only \acp{UAV} are available. \ac{SI} focus on keeping all the essential systems interacting properly and operating with the best performance possible given the adverse conditions. Moreover, \ac{SI} was designed following standards, mainly the \ac{ETSI} \ac{ENI} specifications \cite{ENI-2020}. Regarding the results, one of the experiments indicates that \ac{SI} can improve MC services by splitting neural networks between \ac{UAV} and edge equipment, such that the latency decreases by 29.87\%. In another experiment, it is shown that adequately slicing the resources leads to a decrease in average latency from 66~ms to 34~ms, approximately, when considering the downlink communication between \ac{UAV} and edge.

The remaining of this article is organized as follows. In Section~\ref{sec:background}, we present the standardized approach's background, considering the main initiatives towards \ac{AI/ML} solutions. In Section~\ref{sec:proposal}, we introduce our System Intelligence solution, and in Section~\ref{sec:perf_eval}, we show the experiments with AI for \ac{UAV}-based \ac{SAR}. We discuss the related work in Section~\ref{section:iii} and present final remarks and suggestions for future work in Section~\ref{sec:conclusion}.

\section{Background}\label{sec:background}

In the last years,  telecommunications standardization bodies, such as \ac{ITU-T}, \ac{ETSI}, \ac{3GPP}, as well as alliances between operators and manufacturers as \ac{O-RAN}, published specifications about the design, development, and deploy of \ac{AI}/\ac{ML} for the 5G ecosystem and also its evolution, i.e., \ac{B5G}. Together, these specifications encompass a wide scope, including the core, \ac{MEC}, and \ac{RAN}. Systems using \ac{AI} in 5G/\ac{B5G} networks will surely be based on these specifications. However, most \ac{AI} systems used in 5G are not compliant to standards at the current maturity stage yet. Moreover, before \ac{AI} systems become widespread, grasping all related specifications may be a daunting task. The specifications overlap and have gaps concerning some issues. Initially, this section briefly introduces the architectures defined by these standards, which cover \ac{5GC}, \ac{RAN}, and \ac{MEC}, and also the integration between them. After, the section presents a summary of the main initiatives considering the \ac{AI/ML}'s perspective for the 5G/\ac{B5G} ecosystem. Besides providing a short review of these specifications for the reader's convenience, it is also vital to contextualize the proposed \ac{SI} and indicate what can be used from the standards and what was missing.

The network system is mostly standardized by the \ac{3GPP} 5G, composed of the \ac{5GC} built as a \ac{SBA} \cite{3gpp:rel15nr21.915} and the \ac{NG-RAN}. An \ac{SBA} allows flexible and stateless positioning of virtual environments in the network segments that make up a 5G system. Moreover, this architecture refers to how virtual networks' functions are created and deployed flexibly, widely using the concept of cloud computing to develop, deploy, and manage services. In this context, Releases 15 \cite{3gpp:rel15nr21.915} and 16 \cite{3gpp:rel16nr21.916} introduce several features to \ac{5GC} that are useful in the context of \ac{MC} systems. For example, services can be implemented to expand the capabilities of a \ac{MC} system, decomposing functions with low granularity, making the service light, and having a high capacity for sharing.
%ORIGINAL:
%Mission Critical (MC). For example, services can be implemented and expand MC services to decompose functions with low granularity, make the service light, and have a high capacity for sharing. 

When contrasted with \ac{3GPP} specifications, the \ac{O-RAN} Alliance introduced a complementary set of \ac{NG-RAN} standards, gaining relevant support from the telecommunications industry. \ac{O-RAN} addresses the split of the \ac{gNB} in three parts: (i) \ac{O-RU}, (ii) \ac{O-DU}, and (iii) \ac{CU}. These splits for radio access technologies can be designed, developed, and deployed for the sake of saving costs or dealing with restrictions in energy consumption, such as in \acp{UAV} networks. In this context, \ac{3GPP} and \ac{O-RAN} define a disaggregated \ac{RAN}, composed of multiple \ac{VNF}. Therefore, we use the term \ac{vRAN} throughout the article to emphasize that both \ac{NG-RAN} and \ac{5GC} can be implemented as a collection of \ac{VNF} that must be appropriately placed and chained to accomplish their tasks \cite{ghosh20195g}. 

\begin{figure}[!ht]
\begin{center}
\includegraphics[width=1\textwidth]{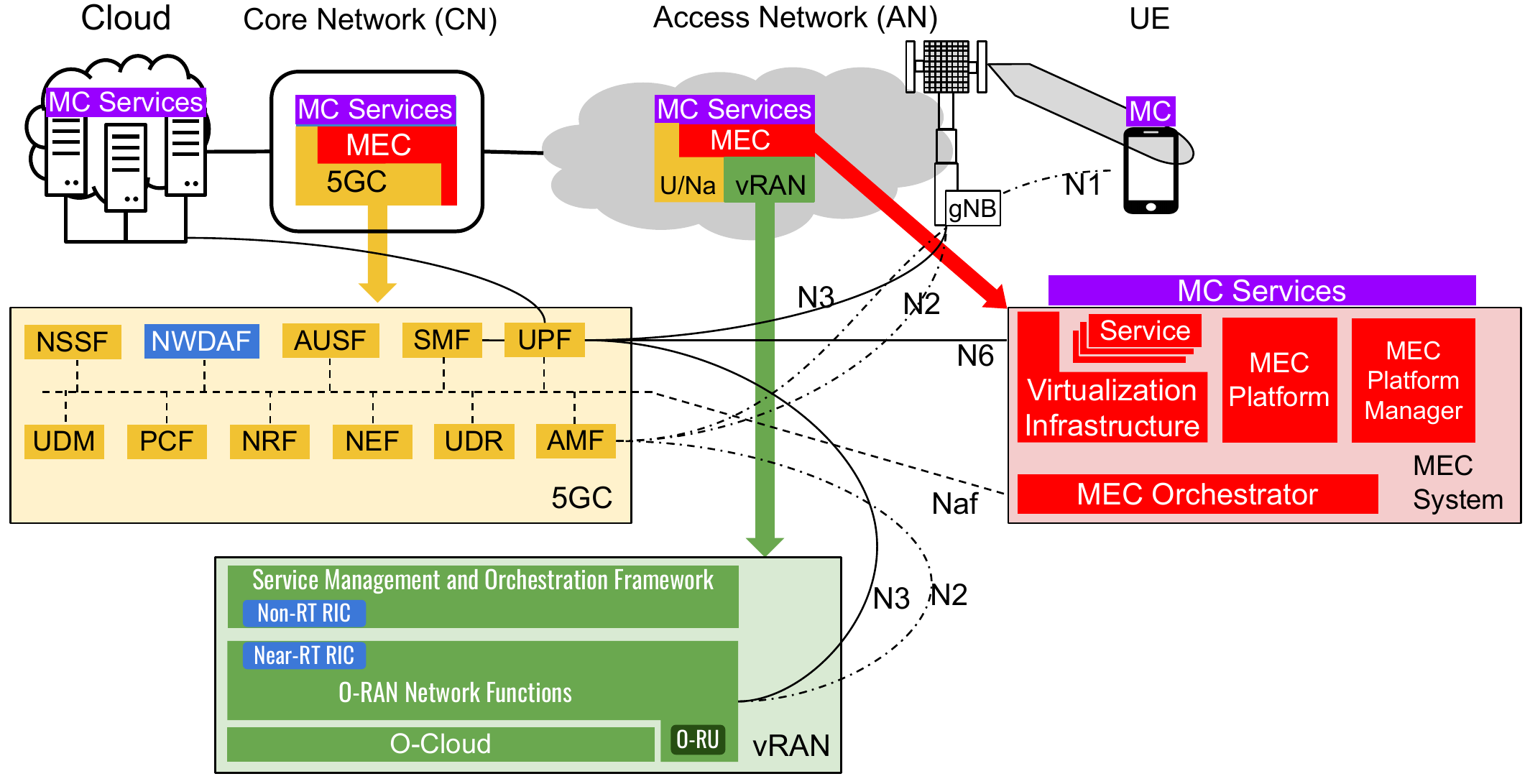}
  \end{center}
\caption{A standardized approach for supporting \ac{MC} services based on the integration of \ac{3GPP} 5G system (mainly core), O-RAN, and \ac{ETSI} \ac{MEC} system. The \ac{AI}/\ac{ML} components already defined in the standards are highlighted in blue.}
\label{fig:MC-support}
\end{figure}

Given the importance of edge computing in \ac{MC} scenarios, \ac{ETSI} efforts in this area are briefly described.
\ac{ETSI} specified a \ac{MEC} system as software-only entities to operate on top of a network edge's virtualization infrastructure \cite{etsi-mec-wp20}. This system consists of a platform to run applications on a particular virtualization infrastructure, a platform manager to handle the specific functionality's management, and an orchestrator, which controls the whole system and services. Moreover, the \ac{MEC} system is aligned with the \ac{SBA} principles, and it was designed to be tightly integrated with \ac{5GC} \cite{etsi-mec-wp28}. Some of the essential \ac{MEC} services depend on this integration, with \acp{API} such as Radio Network Information API, Location API, \ac{UE} Identity API, and Bandwidth Management API.

Figure~\ref{fig:MC-support} illustrates how \ac{MC} services can be supported by a standardized set of systems composed of a \ac{5GC}, \ac{vRAN} according to the O-RAN architecture, and an \ac{ETSI} \ac{MEC} system. All these systems can be seen as a large and complex collection of \acp{VNF} that must adequately be managed and orchestrated to support users' application and services. The \ac{MC} services add the challenge of managing and orchestrating these \acp{VNF} over an infrastructure composed mainly of \acp{UAV} and that can constantly change. Additionally, the figure illustrates how interconnected are multiple systems. In fact, \ac{SBI} \ac{Naf}, and the reference points N1, N2, N3, and N6 are only some of the interconnections among the \ac{5GC}, \ac{vRAN}, and \ac{MEC} system. Finally, the figure highlights (in blue) the components related to \ac{AI}/\ac{ML} already introduced by the standards. 

\acp{UAV} have been used to aid military and rescue operations under challenging areas \cite{azmat2020potential, hossain2020integration}, performing several tasks, such as surveillance, inspection, and mapping \cite{orfanus2016self}. The presence of \acp{UAV} in these scenarios tends to increase, with them being used for even more applications, including disaster management  \cite{li2018uav, qadir2021addressing}, performing \ac{MC} tasks, leveraging the capacity of the new generation of mobile networks. However, it is good to note that MC services provided in disaster scenarios usually need to deal with reduced or even lack of radio coverage \cite{alsaeedy20205g}. For instance, recent hurricanes occurrences, such as Hurricane Maria, resulted in the degradation of the functionality of about 95.6\% and 76.6\% of cellular sites in the affected areas \cite{safety2018homeland}. In this context, the rescue team needs to ensure that its \acp{UAV} will receive the radio resource allocation dynamically during the mission through a network slice, guaranteeing the necessary throughput and latency, for example, video streaming and remote control.

In \ac{MC} scenarios, \ac{5GC}, \ac{vRAN}, and \ac{MEC} must be improved to be operational under the eventual harsh conditions imposed by the environment. In other words, \ac{MC} is challenging enough to require functionalities that are not still incorporated in current specifications. When seeking a solution that supports MC, part of this solution can be obtained from \ac{3GPP} specifications that describe the \ac{5GC} architecture. For example, when considering the \ac{AI} required by \ac{MC} systems, the component called \ac{NWDAF} is the approach of \ac{3GPP} for meeting the AI perspective in the 5G system written in Release 15 \cite{3gpp:rel15nr21.915} and 16 \cite{3gpp:rel16nr21.916}. This component is responsible for collecting several types of information from the network and its users. Any core component and external access can consume the services provided by \ac{NWDAF}. The analytical data produced by \ac{NWDAF} can be used by an \ac{AI} agent that specifies actions in the network context, for example, for UAV-based critical missions. In this case, the \ac{5GC} plays proactively and takes real-time decisions to provide the necessary services to 5G users, and \ac{NWDAF} becomes a central point for analytics in the 5G network.

\begin{figure}[!ht]
\begin{center}
\includegraphics[width=0.9\textwidth]{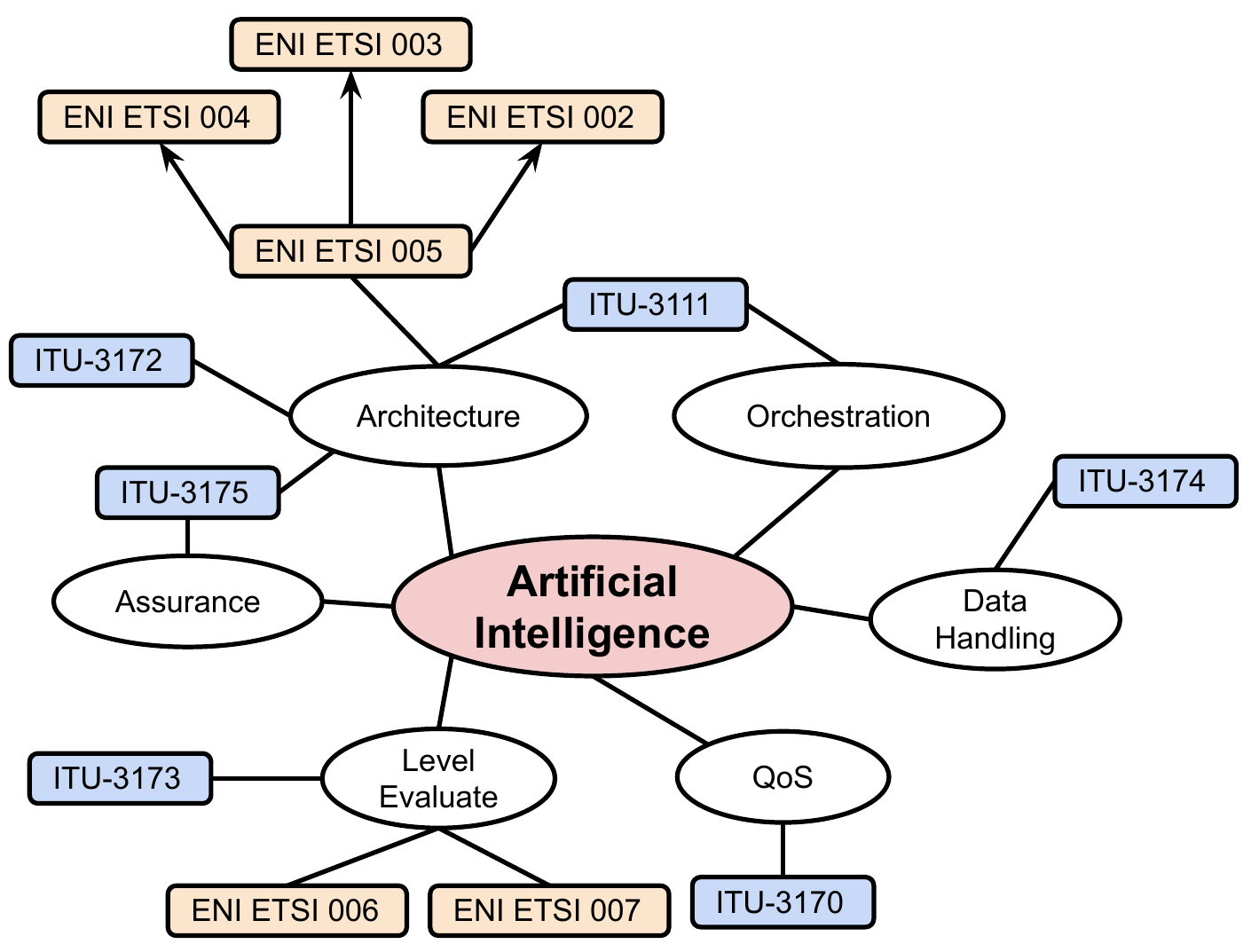}
  \end{center}
\caption{Artificial intelligence standards for communication networks by ITU and ETSI.}
\label{fig:Standardizing}
\end{figure}
 
Complementing \ac{5GC} by \ac{3GPP}, the \ac{O-RAN} architecture defines an \ac{AI} perspective, including the AI-enabled \ac{RIC} for both \ac{non-RT} and \ac{near-RT} \cite{Shafin-2020}. The \ac{non-RT} functions include service and policy management, higher layer procedure optimization, and model training for the \ac{near-RT} \ac{RAN} functionality. The \ac{near-RT} \ac{RIC} is fit with radio resource management and improves operational functions such as seamless handover control, \ac{QoS} management, and connectivity management.

% \ak{And ETSI announced manuscripts exploring the opportunities and new challenges to exploring AI on the 5G ecosystem.} 

% Considering the AI context for improving the automation and decision-making processes, ITU-T presents recommendation sets for applying AI in the telecommunication systems.

The specifications published by \ac{3GPP} and \ac{O-RAN} are fundamentals and applied for using of \ac{AI}/\ac{ML} solution on \ac{5GC} and \ac{vRAN}, respectively. 
In a broader scope, \ac{ITU-T} and \ac{ETSI} published a set of manuscripts to define terminology, requirements, functionalities, and other essential concepts for characterizing \ac{AI} integration into communication networks. We organize the manuscripts from these two standardization bodies considering the \ac{AI} perspective in Figure \ref{fig:Standardizing} and summarize these specifications in Table \ref{Table:Standard}. \ac{ITU-T} introduces six main specifications that cover \ac{AI} in a next-generation network. These recommendations focus on several aspects of the design of an \ac{AI} framework. For example, ITU-3111 presents the network management and orchestration framework, and ITU-3170 shows the requirements for machine learning-based \ac{QoS} assurance for the network. Moreover, ITU-3172 introduces the architectural framework for machine learning, and ITU-3173 discusses the framework for evaluating the intelligence levels of networks. Furthermore, ITU-3174 considers the framework for data handling to enable machine learning, and ITU-3175 covers the functional architecture of machine learning-based quality of service assurance in the networks.

\begin{table}[!ht]
\centering
\caption{AI/ML specifications by ITU-T and ENI ETSI \label{Table:Standard}}
\def\arraystretch{1.2}
\begin{tabular}{l|l}
\hline
\textbf{Specification} & 
\textbf{Title}                       \\ \hline

%%%%%%%%%%%%%%%%%%%%%%%%
\rowcolor[HTML]{C5DEED} ITU-3111 &  Network management and orchestration framework
\\ \hline

%%%%%%%%%%%%%%%%%%%%%%%%
ITU-3170 & Requirements for ML-based quality of service assurance for \\ & the IMT-2020 network
\\ \hline

%%%%%%%%%%%%%%%%%%%%%%%%
\rowcolor[HTML]{C5DEED} ITU-3172 & Architectural framework for ML in future networks including \\
\rowcolor[HTML]{C5DEED} & IMT-2020
\\ \hline

%%%%%%%%%%%%%%%%%%%%%%%%
ITU-3173 & Framework for evaluating intelligence levels of future networks
\\ 
& including IMT-2020
\\ \hline

%%%%%%%%%%%%%%%%%%%%%%%%
\rowcolor[HTML]{C5DEED} ITU-3174 & Framework for data handling to enable ML in future networks
\\ 
\rowcolor[HTML]{C5DEED} & including IMT-2020
\\ \hline

%%%%%%%%%%%%%%%%%%%%%%%%
ITU-3175 & Functional architecture of ML-based QoS assurance for \\  & the IMT-2020 network
\\ \hline

%%%%%%%%%%%%%%%%%%%%%%%%
\rowcolor[HTML]{C5DEED} ENI ETSI 002 & ENI requirements
\\ \hline

%%%%%%%%%%%%%%%%%%%%%%%%
ENI ETSI 003 & Context-Aware Policy Management Gap Analysis 
\\ \hline

%%%%%%%%%%%%%%%%%%%%%%%%
\rowcolor[HTML]{C5DEED} ENI ETSI 004 & Terminology for Main Concepts in ENI
\\ \hline

%%%%%%%%%%%%%%%%%%%%%%%%
ENI ETSI 005 & Experiential Networked Intelligence - System Architecture 
\\ \hline

%%%%%%%%%%%%%%%%%%%%%%%%
\rowcolor[HTML]{C5DEED} ENI ETSI 006 & Proof of Concepts Framework
\\ \hline

%%%%%%%%%%%%%%%%%%%%%%%%
ENI ETSI 007 & Definition of Categories for AI Application to Networks
\\ \hline

\end{tabular}
\end{table}

Figure~\ref{fig:Standardizing} also identifies some key \ac{ETSI} documents. \ac{ENI} \ac{ETSI} 006 and 007 address concepts and definitions of categories for \ac{AI} applications to networks, such as planning and optimization, service provisioning and assurance, data management, operator experience, etc. Moreover, \ac{ENI} \ac{ETSI} 0055 presents a high-level architecture for experiential networked intelligence \cite{ENI-2020}, as well as additional contents showing experimental network requirements (\ac{ENI} \ac{ETSI} 002), context-aware policy management (\ac{ENI} \ac{ETSI} 003), and terminology for experimental networks (\ac{ENI} \ac{ETSI} 004). This architecture is referred to as an \ac{AS} composed of three classes that represent: (i) no AI-based capabilities, (ii) \ac{AI} is not in the control loop, and (iii) \ac{AI} capabilities in its control loop. Additionally, the architecture designs an \ac{API} Broker to serve as a gateway between different systems. 

Even with such a relatively short review, it is clear that the body of recent work aiming at defining how \ac{AI}/\ac{ML} should be used in future networks is considerable. Before some of these recommendations become widely adopted in the industry, it is essential to identify how they can be put together and what remains to be defined. The next section uses some key aspects of these recommendations and describes missing elements to compose an intelligent system suitable for critical missions.

\section{System Intelligence}\label{sec:proposal}

In this section, we describe our proposal in detail, i.e., \ac{SI}, and how it can deal with the challenging scenarios previously introduced. Since adherence to standards is one of our primary concerns, we built \ac{SI} mainly in compliance with \ac{ETSI} \ac{ENI}~\cite{etsi-eni-005}. However, while \ac{ENI} is a general-purpose abstract specification, \ac{SI} is an instantiation of the \ac{ENI} ideas, being focused on \ac{MC} services scenarios and taking into account a complex infrastructure that includes computing devices, \acp{gNB}, and, mainly, \acp{UAV}.  

\begin{figure}[!h] %[htbp] 
 \begin{center}
\includegraphics[width=0.9\textwidth]{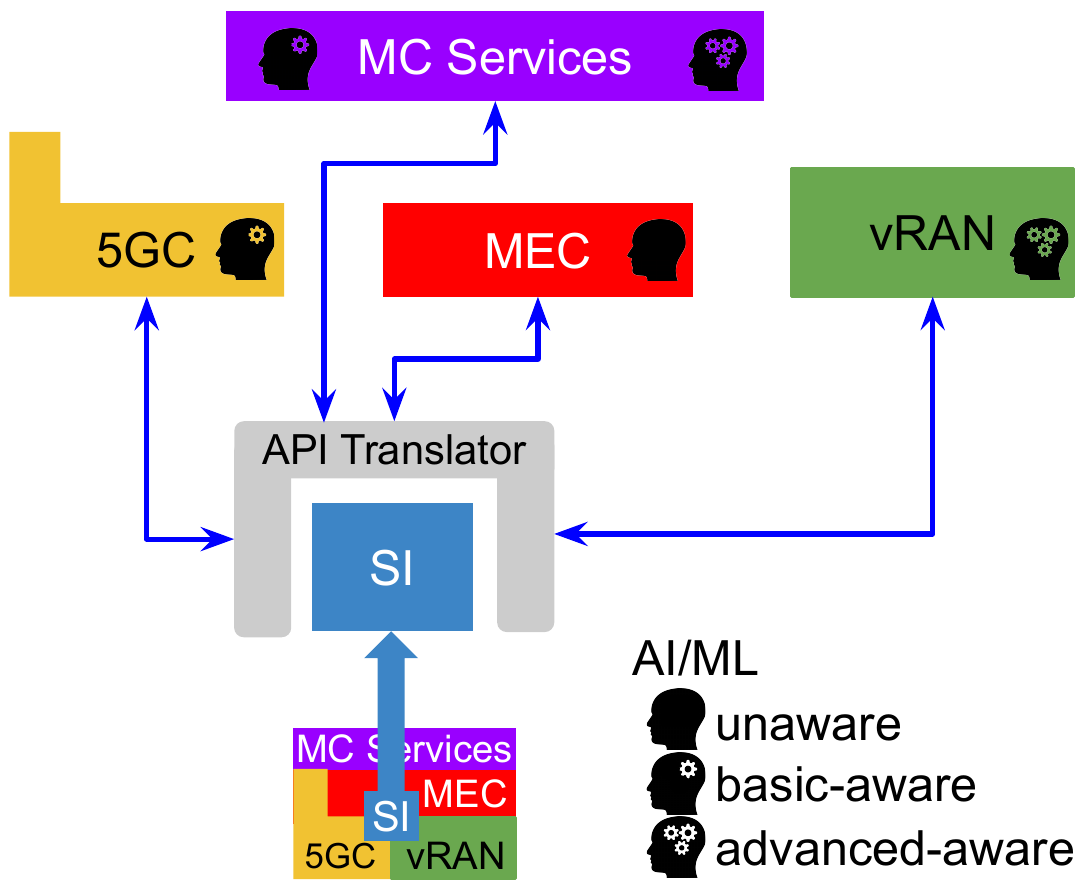}
  \end{center}
%\caption{\ac{SDAR} examples faced by the support of the \ac{EI} system, illustrating the need for \ac{SI}.}
\caption{System Intelligence integration with the main Assisted Systems of the Mission Critical scenarios.}
\label{fig:SI4MC}
\end{figure}

In Figure~\ref{fig:SI4MC}, we illustrate the introduction of \ac{SI} and its integration with the standardized systems previously presented. Similar to the \ac{ENI} specification~\cite{etsi-eni-005}, these systems that are managed by or receive recommendations from \ac{SI} are named \textit{Assisted Systems}. Moreover, as in \cite{etsi-eni-005}, we recognize that each assisted system may present a distinct level in terms of capabilities related to \ac{AI/ML}-based decision-making. However, we adopt a different terminology to categorize the systems, which is more appropriate to the \ac{MC} services context, as described in the following.

We consider the \ac{MEC} System as \textit{AI/ML unaware}, which means that it has no \ac{AI/ML}-based decision-making capabilities. We categorize \ac{5GC} as \textit{AI/ML basic-aware}, since it has some \ac{AI/ML}-based decision-making capabilities, mainly characterized by the \ac{NWDAF} component. Finally, we consider \ac{vRAN}, as defined by \ac{O-RAN}, an \textit{AI/ML advanced-aware} system, which means the existence of sophisticated \ac{AI/ML}-based decision-making capabilities, including an internal \ac{AI} in the control loop. \ac{MC} services are provided by multiple \ac{AI/ML} applications, which can be either basic-aware or advanced-aware. From any category, \ac{SI} only needs a set of \acp{API} that allows it to obtain data from the assisted system (related to its state) and to provide commands or recommendations for how to act, in general, to achieve a specific goal. It is important to highlight that no change is necessary for any assisted system to interact with \ac{SI} appropriately. Actually, not even \acp{API} need any adjustment. As illustrated in Figure~\ref{fig:SI4MC}, a component named \textit{API Translator} is responsible for translating between \acp{API} of \ac{SI} and \acp{API} of the Assisted Systems, if necessary.

\begin{figure*}[!h] %[htbp] 
 \begin{center}
\includegraphics[width=1\textwidth]{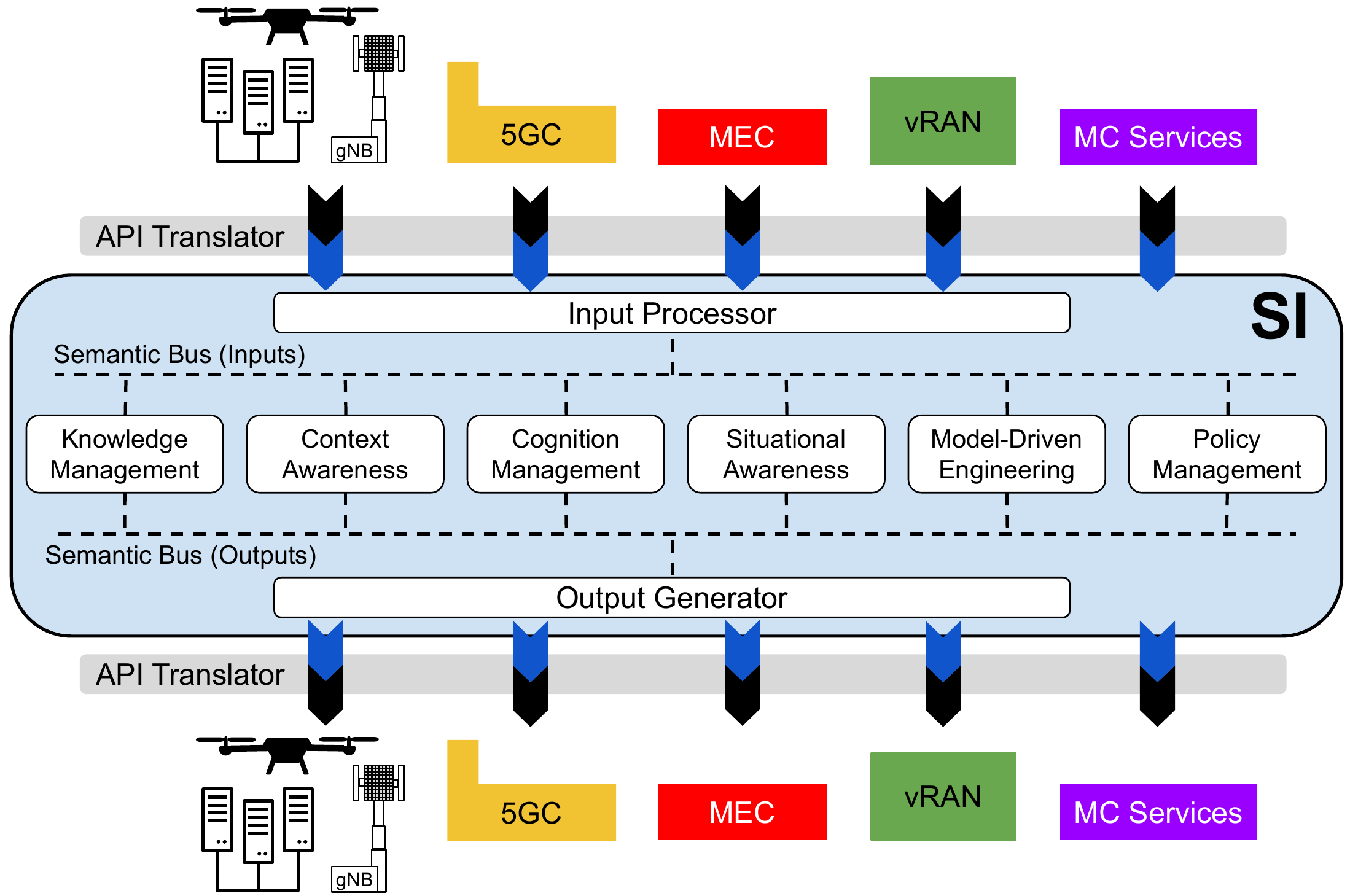}
  \end{center}
%\caption{\ac{SDAR} examples faced by the support of the \ac{EI} system, illustrating the need for \ac{SI}.}
\caption{System Intelligence architecture based on \ac{ETSI} \ac{ENI} architecture.}
\label{fig:SI-arch}
\end{figure*}

The main elements of \ac{SI} are presented in Figure \ref{fig:SI-arch}. The \textit{Input Processor} element handles several types of data from different sources using connection points between the \textit{\ac{API} Translator} and the external systems, e.g., \ac{5GC}, \ac{MEC}, and \ac{vRAN}. These data are normalized and forwarded to the \textit{Semantic Bus}, which connects with the six internal \ac{SI} elements \cite{Shafin-2020}. These internal elements generate results from processing and making decisions based on these standardized data. The results can be new facts or new hypotheses, which can later be converted into actions (by the \textit{Output Generator}) to be applied to the systems. In the following, we describe each of the internal \ac{SI} elements and introduce their deployment considering \ac{MC} services and the infrastructure components, especially \acp{UAV}.

\subsection*{Knowledge Management}

This component defines formalism for representing information and knowledge, enabling the \ac{SI} to analyze, apply, and validate decision-making processes. Knowledge Management works with data and information, using a knowledge representation that defines mechanisms for the set of entities' characteristics and behavior. Moreover, this component enables \ac{SI} to plan actions and determine consequences by \ac{AI/ML} and reasoning to direct action on the set of entities. In this context, this component handles which context and situation information is applied to the raw data, transforming it to information and then knowledge. 

Assuming a \ac{UAV}-based \ac{MC} use case, \acp{UAV} can provide essential data about the context and situation. For example, object detection with \ac{AI/ML} algorithms applied to videos from searching and rescuing areas can help locating victims. The \ac{AI/ML} algorithms needed during the search stage may change from those required for diagnostic and rescue.
%\ak{For instance, assume that \acp{DNN} are used for object detection~\cite{Wang-2020}, but the set of objects that need to be detected change over time.} 
For instance, assume that \acp{DNN} are used for object detection~\cite{Wang-2020}, but the set of objects needs to be detected change over time.
Therefore, \ac{SI} needs to have Knowledge Management to fit different \ac{AI/ML} algorithms loaded at \acp{UAV} for each context and situation. Moreover, \acp{UAV} can be positioned for providing connectivity in remote areas, being configured on-the-fly as repeaters, just amplifying the signals, or as a full \ac{BS} \cite{moradi2019skycore}. In this case,  Knowledge Management using context and situation information represented by estimated interference signals and coverage areas, can assist in the decision-making that regards choosing the better connectivity strategy for each \ac{MC} service.

\subsection*{Context Awareness}

The Context Awareness component enables various data and information to be easily correlated and integrated with the other \ac{SI} components. In this case, this component allows \ac{SI} to provide customized services and resources corresponding to that context. This component's fundamental characteristic is to enable \ac{SI} to adapt its behavior according to changes in the context. For example, the contextual history may be useful for driving policy decisions for current and future interactions. Moreover, context knowledge offers a greater level of reliability and usefulness over the whole systems.

\ac{SI} must store data regarding the network, services, and users, such as the number of users and the bandwidth requested along with an \ac{MC}. For example, it is possible to ensure that applications that benefit from low latency are prioritized over others with less importance. The data considered by Context Awareness is mandatory for instance, in \ac{SAR} scenario that deals with the transmission of \ac{CS} to inform \acp{UAV} about the flight plan for an \ac{MC} \cite{3gpp:rel17UAV, Pham-2020} task, such as person identification or supply delivery. 

\subsection*{Cognition Management}

This management allows \ac{SI} to understand data and information input in the system, i.e., defining how these were produced. In this case, the component provides four functions: (i) perform inferences to generate new knowledge, (ii) change existing knowledge, (iii) use raw data and historical data to learn what is happening in a distinct context and situation, e.g., why the data were generated, and which components could be affected, and (iv) determine new actions to guarantee the goals of \ac{SI}. Finally, the Cognition Management component uses these four functions to validate and generate new knowledge. 

An \ac{MC} service usually operates in dynamic scenarios regarding
%, e.g., to search and rescue, considering the
computation-communication trade-offs between the \acp{UAV} and network infrastructure. In this context, \ac{SI} could provide Cognition Management to choose, change, and deploy placement strategies for the computational vision and \acp{VNF}, evaluating their effects on CPU and RAM usage,
%, score transmission bit rate over \acp{UAV} and network infrastructures. 
%AK: nao definimos ainda os scores
Moreover, all the changes over the configuration of systems and flight plans must be handled through their history to generate new knowledge.

\subsection*{Situational Awareness}

This component enables \ac{SI} to understand what happened and how it influences the \ac{SI} goals. This component works to know how and why the current situation presents such results. \ac{SI} observes various situations evolving, examining them for patterns within each condition and between different cases. This observation process includes five actions: (i) collecting data, (ii) understanding the significance of this data, (iii) determining what to perform in response to a given event, (iv) making a decision, and (v) evaluating these actions. In this way, it enables the application of context and policies to a distinct situation using inference and historical data to learn what is happening in one specific context, why, and what should be done in response to it. It is essential to highlight the difference between Context Awareness and Situational Awareness. The first one describes the state and environment in which an entity exists. The second one incorporates contextual information and other inputs to understand the meaning of data and behavior of the entire assisted system and its operational environment. 

One of \ac{UAV}'s essential applications in \ac{MC} service is to provide a quick evaluation of the situation in \ac{SAR} areas. \acp{UAV} promote versatility, fast response time, and capacity to support several services. Therefore, all this requires ability to learn about various evolving situations and examine patterns within each condition among different \ac{MC} services. In this case, these services can allow Situational Awareness to keep predicting the situation's progression and how it affects the goals to achieve the \ac{MC} task.

\subsection*{Model-Driven Engineering}

Model-Driven Engineering is an approach to software development where models are used to understand, design, implement, deploy, operate, maintain, and modify software systems. This component focuses on business logic, using an abstract methodology. Therefore, this model supports three essential purposes: (i) to ensure that several data models used in \ac{SI} maintain a consistent definition and understanding of concepts, (ii) to enable different policies at various levels of abstraction to communicate with each other using a common vocabulary and data dictionary, and (iii) to develop from a specification of policy to its implementation.

The integration of several standards for supporting \ac{MC} services, such as \ac{3GPP} \ac{5GC}, \ac{O-RAN}, and \ac{ETSI} \ac{MEC}, demands software development models to provide abstraction levels and a shared vocabulary between external systems and \ac{SI}. It is essential to define data models and \acp{API} for supporting this integration among the standards. Moreover, another crucial feature of model-driven is to design different \ac{APN} to direct users to various service networks, for example, to guarantee the connectivity of \acp{UAV}, \ac{vRAN}, \ac{MEC}, \ac{5GC}, and \ac{SI}, to offer the better \ac{MC} service.

\subsection*{Policy Management}

Policy Management provides uniform and intuitive mechanisms for providing consistent recommendations and commands to ensure the scalable decisions directing \ac{SI} behavior. Three types of policies can be used in \ac{SI}. The Imperative Policy uses statements to change the state of a set of targeted objects explicitly. Declarative Policy works based on ideas to describe what needs to be done without defining how to execute this task. The Intent Policy applies statements from a restricted natural language to express the policy's goals, but not how to accomplish these goals. \ac{SI} can handle any combination of these policies to define recommendations and commands to support and manage the system.

An \ac{MC} service is fundamental for controlling the network's behavior, applying security and control rules related to \ac{UAV}'s session management, mainly for functionalities associated with \ac{vRAN}. This component should provide a mobility policy to add the control of access restrictions to \ac{MC} services in a given area. Moreover, the policy should include the management of topics associated with priority access to the channel of given \ac{UAV} to others detriment. Furthermore, this management should provide metrics related to \ac{QoS} and information regarding the data flow, which is obtained by regularly monitoring events, for example, considering the transmission bit rate on the partitioning strategy for the computational vision among \acp{UAV}, \ac{vRAN}, \ac{MEC}, and \ac{5GC}.

The next section presents experiments that aim at assessing specific aspects of using \ac{SI} in \ac{MC} scenarios. The description of a \ac{SI} system working in a closed-loop, retrieving data, and imposing actions are out of this article's scope. However, the following section presents evidence using an actual testbed on how \ac{MC} services performance can be improved via \ac{SI}.

\section{Experiments with AI for \ac{UAV}-Based SAR}
\label{sec:perf_eval}
% Use case? 

% Cleverson:
% ambição é estabelecermos duas fatias (segmentos), uma para controle e outro dado, fim a fim (ou seja, não só a \ac{RAN}), garantindo isolamento entre as duas fatias.
% Fato: quando a rede de comunicação inclui "Internet", não há como garantir isolamento das fatias.
% Daí, nos cenários onde temos "cloud" processing, podemos desistir de controlar as duas fatias, mas no edge isso seria viável.
% Assim, iniciemos com o caso do "edge" processing, para garantir funcionamento em caso simples.

This section describes \ac{SAR} experiments as a use case to make more concrete the discussion about the benefits of having \ac{SI} in \ac{MC} services. The goal of the experiments is twofold: (i) provide concrete examples of how \ac{SI} orchestrating communications and \ac{AI}/\ac{ML} applications can positively impact \ac{MC} services and (ii) demonstrate how open-source software and low-cost off-the-shelf equipment can be put together to assess essential issues related to \ac{AI}/\ac{ML} in \ac{5G}/\ac{B5G} \ac{MC} scenarios.

%With respect to communications, it is assumed a
The \ac{5G}/\ac{B5G} network is where \ac{SI} can orchestrate tasks such as the placement of \acp{VNF} in different network elements. \acp{UAV}, for instance, can behave as \acp{UE} or \acp{BS}, under the control of \ac{SI}. Regarding \ac{AI}/\ac{ML}, \acp{UAV} fly around a disaster area with equipment for performing computational vision (more specifically, object detection via \acp{DNN}). \ac{SI} has the flexibility of distributing \ac{AI}/\ac{ML} processing among equipment in the cloud, edge, and \acp{UAV} themselves. This flexibility allows \ac{SI} to trade-off energy consumption and latency, for instance. Motivated by a \ac{SAR} situation in which the objects to be detected change along with the mission, it is assumed that \ac{SI} can partition (split) the layers of a \ac{DNN} into subsets and allocate distinct equipment to execute each subset of layers. For instance, the first layers of a convolutional \ac{DNN}, which detect simpler features, can be kept fixed and executed by \acp{UAV} (acting as a \ac{UE}). The last layers, which are specialized to the objects of interest, can be implemented by \ac{MEC} (in \ac{vRAN} or in the core) and continuously adapted according to the time evolution of the \ac{SAR} mission.

% In the conceived scenario, \ac{UAV}s fly around a disaster area with equipment for performing both computational vision (e.\,g., object detection via \ac{DNN}s) and \acp{VNF} (e.\,g., such that the \ac{UAV} performs the role of a base station). 
% There is flexibility to distribute processing among equipment in the cloud, edge and \ac{UAV}s themselves. This allows the \ac{SI} to trade off energy consumption and latency, for instance. Similarly, the flexibility in \ac{VNF} placement allows the network to mitigate challenges in connectivity, such as loss of communication between edge and cloud.

The communication network used in the experiments is implemented in a testbed made available to promote reproducible results.\footnote{https://github.com/lasseufpa/connected-ai-testbed} This testbed is based on the \ac{OAI} software for \ac{vRAN} deployment and the free5GC software to implement a Non-Standalone \ac{5GC}. These two elements are containerized with Docker, with automated deployment via Kubernetes. A summary of the hardware used to execute the main software components is presented in Table~\ref{tab:testbed_details}. The testbed employs Mininet to implement both the fronthaul and backhaul topology. Mininet allows emulating switches and routers in the transport network. In this case, the connection between \ac{vRAN} and \ac{5GC} can have different network topologies and behaviors, e.g., different link latencies, bandwidths, and packet loss rates. Therefore, Mininet allows conveniently emulating edge or cloud scenarios along with the experiments, reproducing characteristics found in real scenarios. The automated Kubernetes cluster orchestration complements this flexibility. Moreover, this automation makes it easier to place \ac{5GC} and \ac{gNB} \acp{VNF} (implemented as containers) according to the mobile network scenario defined in each experiment.

%SAR application (e.g., vehicles and people localization) 
We implemented object detection with OpenCV and Pytorch. The layers of the adopted \ac{DNN} were partitioned into two, and these subsets were executed by \ac{UAV} (the first subset of layers) and a cloud or edge equipment (the second subset of layers). The \ac{DNN} layers run by \acp{UAV} were processed by an NVIDIA Jetson Nano, as indicated in Table~\ref{tab:testbed_details}. This board has a quad-core ARM A57@1.43 GHz CPU, with 4 GB 64-bit LPDDR4 memory and a 128-core NVIDIA Maxwell GPU. %Maxwell™ GPU.
Therefore, Jetson Nano represents a situation in which the \ac{UAV} hardware may not execute the full \ac{DNN} in real-time or the last layers change along with the mission.

%The hardware is composed of computers with Intel Core i5-7500 CPU@3.4 GHz processor, running Ubuntu 18.04 LTS as the operating system. These computers host two entities: \ac{5GC} and 5G base station (gNB).

% To represent a \ac{UAV} acting as part of the network (UAV-BS), an Intel Minnowboard Turbot with a quad-core Intel Atom® SoC E3845 (4 x 1.91 GHz, 2 MB cache) CPU and 2 GB DDR3L of onboard RAM was utilized given its compatibility with the \ac{OAI} software due to having Intel architecture.

% \textcolor{blue}{Mostrar em uma tabela as configurações}

% Information for experiments reproduction

% \begin{table}[!ht]
% \centering
% \caption{Testbed functionalities with associated hardware and software}
% \label{tab:testbed_details}
% \def\arraystretch{1.2}
% \begin{tabular}{l|c|c|c}
% \hline
% \textbf{Function} & 
% \textbf{CPU} & 
% \textbf{RAM} &
% \textbf{Software} \\
% \hline

% %%%%%%%%%%%%%%%%%%%%%%%%
% \rowcolor[HTML]{C5DEED}CORE (UPF, AMF, HSS, SMF, PCRF)& 
% i5-7500 & 
% % 8GB DDR4 &
% 8~GB &
% Free5GC \\
% \hline
% %%%%%%%%%%%%%%%%%%%%%%%%
% \ac{RAN} (RCC) & 
% i5-7500 & 
% % 8GB DDR4 &
% 8~GB &
% OpenAirInterface \\
% \hline
% %%%%%%%%%%%%%%%%%%%%%%%%
% \rowcolor[HTML]{C5DEED}RAN (RRH)& 
% i5-7500 & 
% 8~GB &
% OpenAirInterface \\
% \hline%%%%%%%%%%%%%%%%%%%%%%%%
% Backhaul & 
% i5-7500 & 
% 8~GB &
% Mininet \\
% \hline
% % %%%%%%%%%%%%%%%%%%%%%%%%
% % Turbot. & 
% % E3845 & 
% % 2GB DDR3L &
% % \ac{vRAN}or5GC &
% % \\ \hline
% %%%%%%%%%%%%%%%%%%%%%%%%
% \rowcolor[HTML]{C5DEED} \ac{UAV} (Jetson Nano)& 
% A57 & 
% 4~GB &
% Pytorch, OpenCV \\
% \hline
% \end{tabular}
% \end{table}

\begin{table}[!ht]
\centering
\caption{Testbed functionalities with associated hardware and software}
\label{tab:testbed_details}
\def\arraystretch{1.2}
\begin{tabular}{l|c|c|c}
\hline
\textbf{Function} & 
\textbf{CPU} & 
\textbf{RAM} &
\textbf{Software} \\
\hline

%%%%%%%%%%%%%%%%%%%%%%%%
\rowcolor[HTML]{C5DEED} UPF, AMF, HSS, SMF, PCRF (5GC) & 
i5-7500 & 
% 8GB DDR4 &
8~GB &
Free5GC \\ \hline
%%%%%%%%%%%%%%%%%%%%%%%%
%\rowcolor[HTML]{C5DEED} PCRF (5GC) & 
%& 
% 8GB DDR4 &
%&
%\\
%\hline
%%%%%%%%%%%%%%%%%%%%%%%%
RCC (vRAN) & 
i5-7500 & 
% 8GB DDR4 &
8~GB &
OpenAirInterface \\
\hline
%%%%%%%%%%%%%%%%%%%%%%%%
\rowcolor[HTML]{C5DEED} RRH (vRAN)& 
i5-7500 & 
8~GB &
OpenAirInterface \\
\hline%%%%%%%%%%%%%%%%%%%%%%%%
Backhaul & 
i5-7500 & 
8~GB &
Mininet \\
\hline
% %%%%%%%%%%%%%%%%%%%%%%%%
% Turbot. & 
% E3845 & 
% 2GB DDR3L &
% \ac{vRAN}or5GC &
% \\ \hline
%%%%%%%%%%%%%%%%%%%%%%%%
\rowcolor[HTML]{C5DEED} \ac{UAV} (Jetson Nano)& 
A57 & 
4~GB &
Pytorch, OpenCV \\
\hline
\end{tabular}
\end{table}

We devised two sets of experiments. The first concerns evaluating the impact of having \ac{SI} splitting the execution of a \ac{DNN} between a \ac{UAV} (as \ac{UE}) and \ac{MEC} (in \ac{vRAN}, i.e., in edge). The second set of experiments regards \ac{vRAN} slicing and \ac{VNF} placement, which we assume to be also orchestrated by \ac{SI}. In both cases, \ac{SI} is not fully implemented in a closed-loop. Instead of automatically acting according to its inputs, some configurations regarding \ac{AI/ML} and \ac{VNF} placement are manually defined and interpreted as \ac{SI} actions. This configuration simplifies the experiments and their description. A fully working \ac{SI} is out of this article's scope and will be described in future work. We describe the two sets of experiments and their results in the following.
%The next paragraphs describe the two experiments.

%a neural network classification performance variation in detecting people given different splits, useful for search and rescue situations. The second one deals with the effects that \ac{RAN} slicing and network functions placement have on the latency and throughput parameters of a certain \ac{UE}. These manipulations intend to grant priority to a device in order to guarantee or improve the quality of a given service. In this case, the device being prioritized will be \ac{UAV} executing the SAR application. 
%The following experiments explore the benefits of being able to distribute the processing of AI and network functions between the available computational resources by exploring two scenarios:

\subsection{Distributed \ac{AI/ML}}
\label{subsec:split_ai}

Factors such as computational resources, energy consumption, and latency must be considered for real-time image understanding through AI/ML techniques when UAVs capture visual information in MCs. These aspects should be addressed by analyzing the computational cost usage through splitting the processing between the UAV and other processing units.
For this set of experiments, an SSD-VGG16 \ac{DNN} was trained to detect four classes: person, car, bicycle, and motorcycle. The adopted dataset was UFPark, described by Nascimento et al.~\cite{ingrid2020ufpark}, which consists of videos recorded in a campus parking lot. When datasets with videos recorded from \acp{UAV} are available, it will be interesting to compare the performance with models trained with videos obtained from fixed cameras, as adopted in this work.

\begin{figure}[!htb]
    \begin{center}
    \includegraphics[width=1\textwidth]{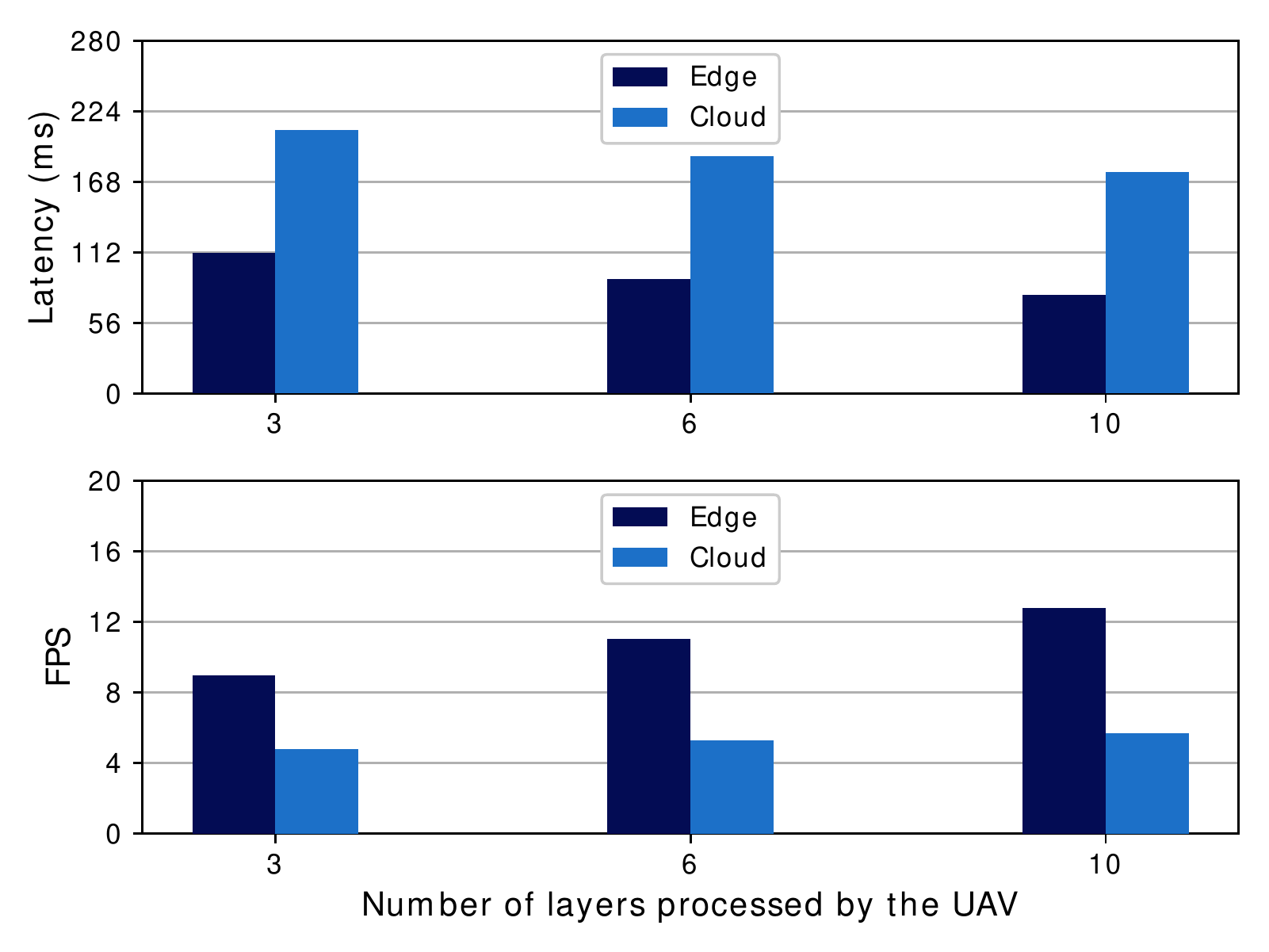}
    \end{center}
    \caption{Performance of \ac{AI/ML} application using different splits of a \ac{DNN} between \ac{UAV} and a terrestrial equipment at  edge or cloud.}
    \label{fig:split_cloud_edge}
\end{figure}

The original video resolution was subsampled to  300$\times$300 pixels with three color channels, and the \ac{DNN} input is a frame of this video. The trained \ac{DNN} had 40 layers and $24.15 \times 10^6$ parameters (weights), each represented with 32 bits. \ac{DNN} was split into two subsets of layers according to three configurations. In the first configuration, \ac{UAV} executed from the first to the third layer. Therefore, the quantized scores (activation values) of this third layer were sent through the network, and a cloud or edge equipment executed the remaining 37 layers. The other two configurations adopted splits after the sixth and tenth layers, respectively.
Figure~\ref{fig:split_cloud_edge} depicts the results for these three configurations concerning the average number of \ac{FPS} (bottom) and the latency (top), considering the time to execute both subsets of \ac{DNN} layers and the time to transmit the data in the uplink from \ac{UAV} to the edge or cloud equipment.
 
%The \textit{max-pooling} reduces the dimension of its input tensor and is an  appropriate layer to perform partitioning: fewer scores need to be transmitted using the \textit{max-pooling} layer output than its input. 

% \begin{figure}[!htb]
%     \begin{center}
%     \includegraphics[width=.45\textwidth]{./figures/edge_lat_fps_fantasy.png}
%     \end{center}
%     \caption{[fantasy values].}
%     \label{fig:edge}
% \end{figure}

The results in Figure~\ref{fig:split_cloud_edge} explore three different split configurations in both edge and cloud scenarios, and identify that, in this case, it is beneficial to have 10 \ac{DNN} layers processed by \ac{UAV}. In contrast, the remaining 30 layers are processed at the edge or cloud. These scenarios need to assess \ac{QoS} measures and adapt the applications running through the transport network indicates that \ac{SI} can improve \ac{MC} services by properly orchestrating communication aspects and also the \ac{AI/ML} applications.

% BACKUP
% According to the results in Fig.~\ref{fig:split_cloud_edge}, it is beneficial to have 10 \ac{DNN} layers processed by \ac{UAV} while the remaining 30 layers are processed at the edge or cloud. This process implies a reduction of latency and an increase in the achievable \ac{FPS}. Processing at the edge instead of at the cloud avoids transmission through the transport network. This reduction allows not only reducing latency but also doubling the FPS that can be achieved by the overall system, as presented in Fig.~\ref{fig:split_cloud_edge}. In this case, executing the whole \ac{DNN} with the Jetson Nano board in the \ac{UAV} leads to unsatisfactory performance. This indicates that \ac{SI} can improve MC services, by properly orchestrating not only communication aspects, but also the \ac{AI/ML} applications.

\subsection{RAN slicing and \ac{VNF} placement}
\label{subsec:placement_slice}

%\begin{figure*}[!ht]
%\centering
%\begin{minipage}[b]{1\textwidth}
% \includegraphics[width=\textwidth]{./figures/throughput_slice.eps}
%    \caption{Throughput in \ac{RAN} slicing experiment.}
%    \label{fig:throughput}
%\end{minipage}\qquad
%\begin{minipage}[b]{1\textwidth}
%  \includegraphics[width=\textwidth]{./figures/latency_slice.eps}
%    \caption{Latency when using \ac{RAN} slicing and \ac{VNF} placement.}
%    \label{fig:latency}
%\end{minipage}
%\end{figure*}

%missions, it is important to assure that the network provides the needed resources to SARs applications to guarantee missions completeness.

\ac{SAR} and other \ac{MC} scenarios require communication networks that maximize the probability of successful mission completion. As indicated by the previous experiments, orchestrating \ac{AI}/\ac{ML} applications can bring significant advantages in realistic situations. Therefore, we now illustrate how \ac{SI} can optimize the network for improved robustness to the \ac{MC} services. In this set of experiments, the \ac{SI} system is assumed to orchestrate \acp{VNF} and manage radio resources to attend two distinct sets of requirements,
%In a \ac{RAN} context, it was executed an experiment demonstrating how \ac{RAN} slicing applied in the radio resource management can be applied to assure required throughput to prioritary applications. 
imposed by applications running on \ac{UAV} (as \ac{UE}) and 
%The experiment considers one \ac{UAV} and 
three traditional \ac{UE}s (e.g., smartphones), respectively. These four devices are connected to a \ac{BS} using \ac{LTE} with a bandwidth of 5~MHz for downlink, which corresponds to a total rate of approximately 18~Mbps to be distributed among the connected devices.

\begin{figure}[!h]
    \begin{center}
    \includegraphics[width=1\textwidth]{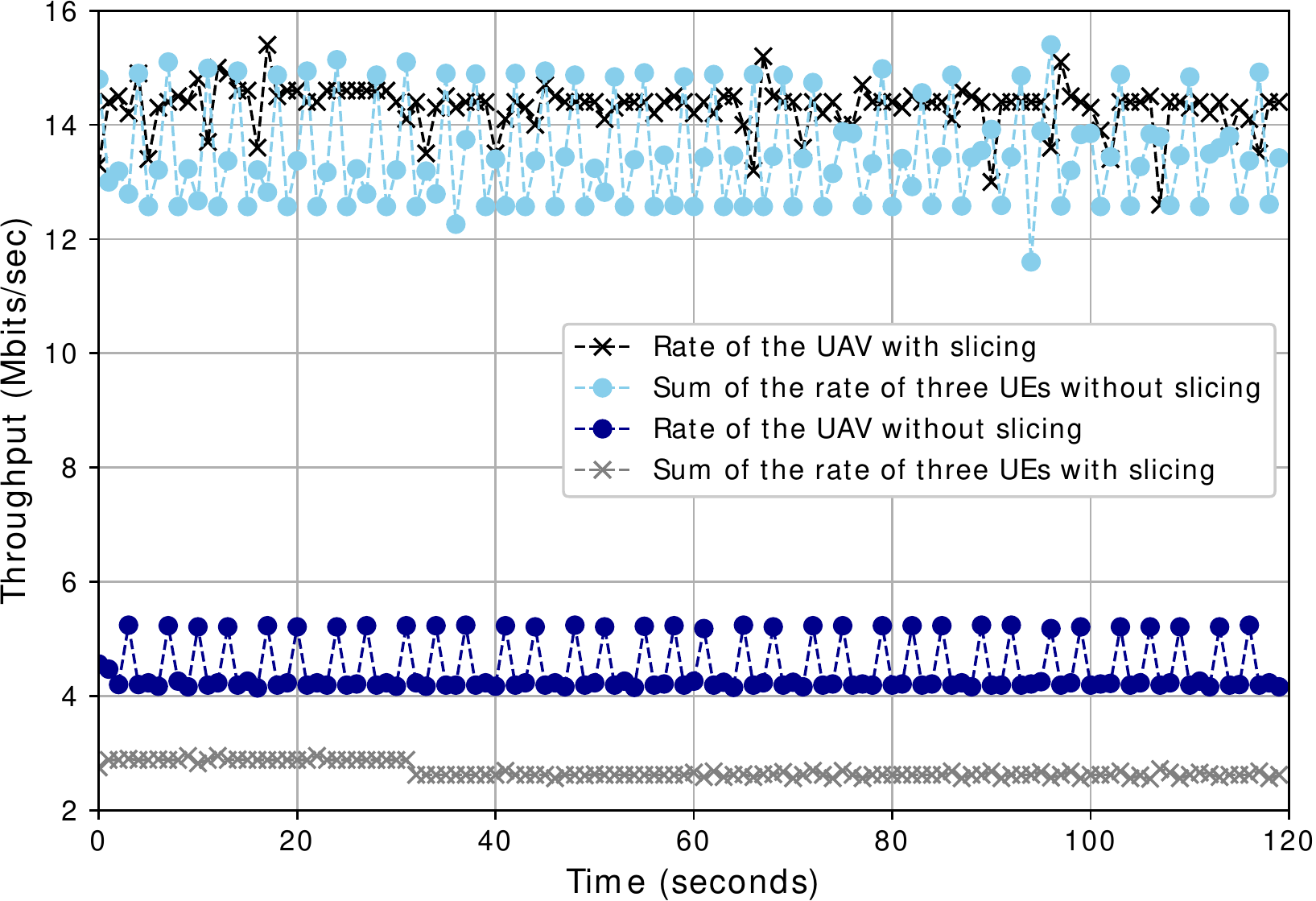}
    \end{center}
    \caption{Throughput in \ac{RAN} slicing experiment.}
    \label{fig:throughput}
\end{figure}

\ac{UAV} executes an MC service for the SAR mission that requires a constant bit rate of 13~Mbps. The other \acp{UE} are requesting 5~Mbps each, for applications that are not critical. We contrast two situations: (a) \ac{SI} imposes a strategy based on \ac{vRAN} slicing to protect the \ac{UAV} communication versus (b) fair scheduling without such \ac{vRAN} slicing. Figure~\ref{fig:throughput} shows these two situations via the throughput values for \ac{UAV} and the three traditional \acp{UE}' accumulated rates. Moreover, the \ac{UAV} application does not reach the required 13~Mbps, considering the scenario without \ac{vRAN} slicing (circle markers). %The base station is using a fair scheduling which tries to provide the same bandwidth to all devices. 
When considering the \ac{SI}-driven scenario using \ac{vRAN} slicing (x markers), the slice exclusively created for \ac{UAV} can guarantee the 13~Mbps target. In this case, the three traditional \acp{UE} only have access to the remaining radio resources, which are equally distributed among them. 

%Therefore, \ac{SI} can protect the radio resources of prioritary applications, as used SAR missions, using \ac{RAN} slicing to provide dedicated radio resources.

\begin{figure}[!h]
    \begin{center}
    \includegraphics[width=1\textwidth]{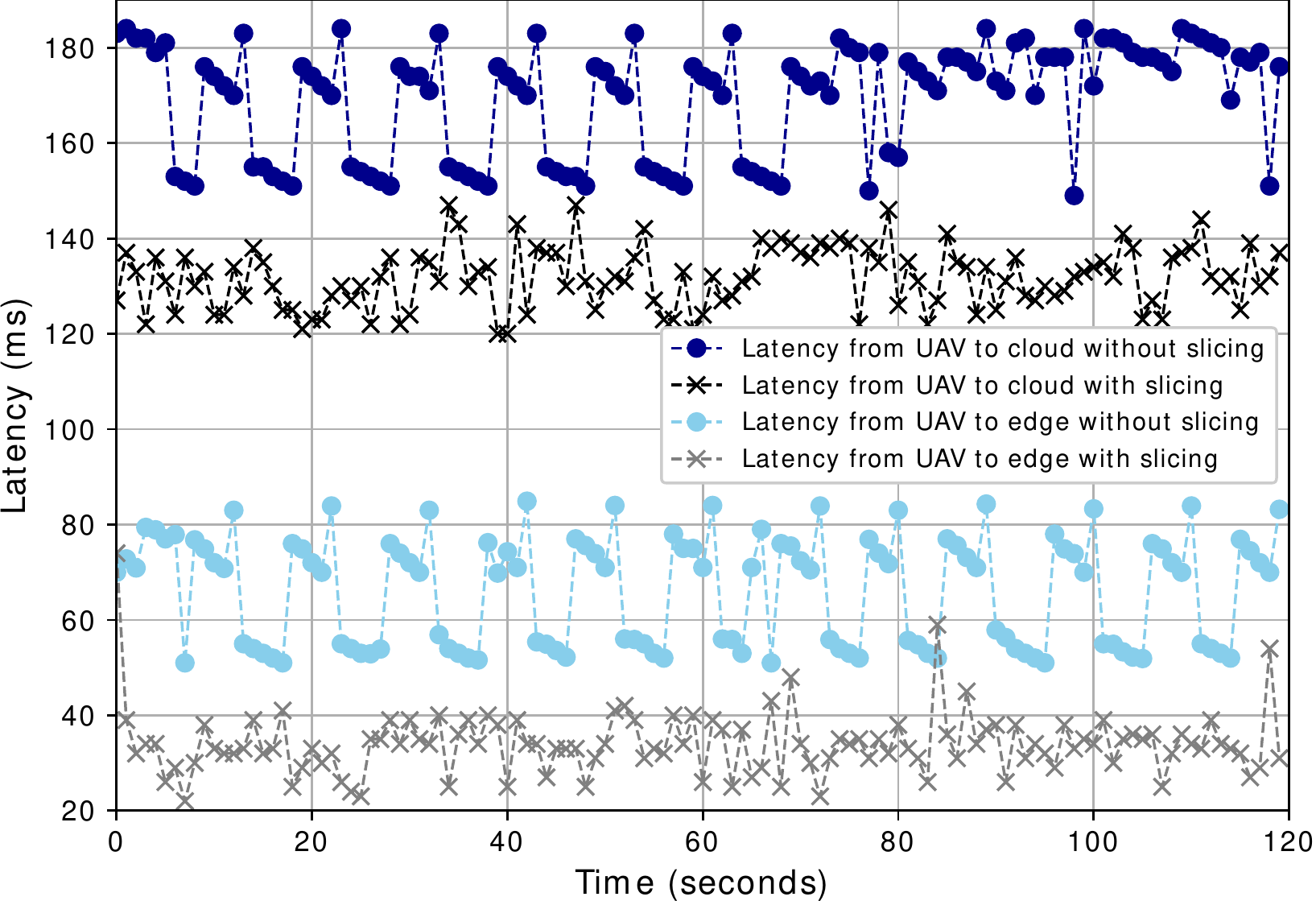}
    \end{center}
    \caption{Latency when using \ac{RAN} slicing and \ac{VNF} placement.}
    \label{fig:latency}
\end{figure}

In Figure~\ref{fig:latency}, the curves contrast the results with and without slicing but concerning latency between \ac{UAV} and equipment at the edge or cloud that is serving \ac{UAV} data. This scenario considers that the same devices of the previous \ac{vRAN} slicing experiment are connected to the network and request the same throughput. 
%Besides the throughput guarantee, other important aspect to SAR missions is the network latency values since real time applications requires low latency. The testbed created two different scenarios where applications used by \ac{UAV} can be both located at Edge or Cloud. 
The application server and \ac{5GC} (with \acp{VNF}, such as UPF and AMF) are either located at the edge or cloud. For emulating the transport network in the cloud scenario, Mininet was used to impose a backhaul topology with a total latency of 100~ms.
%to represent a communication with a distant application located at the cloud.
Moreover, the curves in Figure~\ref{fig:latency} show the effect of both actions: \ac{vRAN} slicing and \ac{VNF} placement. \ac{VNF} allocations on edge provide better latency values than allocating at the cloud. Furthermore, \ac{vRAN} slicing reduces the latency value experienced by \ac{UAV} in both edge and cloud scenarios. 
%AK: o abaixo sugere que conseguimos dizer
%o quao necessario foi, e nao damos conta disso
%Overall, these results indicate how necessary the \ac{SI} action is to guarantee \ac{SAR} demands on throughput and latency.
%AK: Atenuando:
Overall, the results indicate the impact of \ac{SI} actions in guaranteeing \ac{SAR} demands related to throughput and latency.

%Considering the scenario without slice, it is perceived that the devices are competing for the same resources which provocates a bigger latency value and a variation similar to experienced in the throughput experiment. 

% \begin{figure}[!t]
%     \begin{center}
%     \includegraphics[width=.48\textwidth]{./figures/throughput_slice.eps}
%     % \includegraphics[trim={.5cm .5cm .5cm .5cm}, clip, scale=.6]{./figures/rev2_throughput_slice.eps}
%     \end{center}
%     \caption{Throughput in \ac{RAN} slicing experiment.}
%     \label{fig:throughput}
% \end{figure}
% % [trim={0 2cm 0 0}, clip,scale=0.25]
% \begin{figure}[!t]
%     \begin{center}
%     \includegraphics[width=.48\textwidth]{./figures/latency_slice.eps}
%     % \includegraphics[trim={.5cm .5cm .5cm .5cm}, clip, scale=.2]{./figures/rev2_latency_slice.eps}
%     \end{center}
%     \caption{Latency when using \ac{RAN} slicing and \ac{VNF} placement.}
%     \label{fig:latency}
%\end{figure}

\section{Related Work Regarding \ac{UAV}-Based \ac{MC}}\label{section:iii}

Previously, mainly in Section~\ref{sec:background}, we reviewed existing academic articles and standards to contextualize the proposed \ac{SI}. In this section, we complement the literature overview by focusing on a non-exhaustive list of recent manuscripts that discuss integrating the main elements of \ac{AI/ML}-based \ac{MC} services, eventually using \acp{UAV}. As discussed, \ac{MC} applications (e.g., \ac{SAR}) must be supported by telecommunication infrastructures, considering many elements of \ac{vRAN}, edge, core, and even cloud \cite{CENTAURO-19}. In this context, one can observe the great attention of academia and industry 
%on critical mission applications. For example,
for the utilization of \acp{UAV} in \ac{SAR}~\cite{SHULE-2020}. The following paragraphs provide information on how these elements can be put together and help understand the proposed \ac{SI}' relations with previous works.
%Moreover, we view the use of \ac{AI} to adapt the networks dynamically to provide services necessary for critical mission applications. In a not exhaustive way, this section aims to present the state-of-the-art about the use of \ac{AI/ML} in telecommunication infrastructure to offer \ac{MC} services such as \ac{SAR}.

Besides the mentioned communication infrastructure elements, many \ac{MC} applications rely on computational vision \cite{Cazzato-2020}. 
The performance of \ac{MC} applications such as \ac{SAR} can be largely improved by considering \ac{AI/ML}, edge, and multi-\ac{UAV} technologies together~\cite{CENTAURO-19}. Aligned with this perspective, two recent surveys were published \cite{Dianlei-2020, Wang-2020}. Xu et al. \cite{Dianlei-2020} discussed concepts that allow the distinction between Edge Intelligence and Intelligent Edge. For the authors, Edge Intelligence focuses on intelligent applications in the edge environment with edge computing assistance and protection of users’ privacy. On the other hand, Intelligent Edge aims to solve edge computing problems using \ac{AI} solutions, e.g., resource allocation optimization. Moreover, Wang et al. \cite{Wang-2020} emphasize that Edge Intelligence seeks to facilitate \ac{DL} services via edge computing. Furthermore, \ac{DL} can be integrated into edge computing frameworks to build an Intelligent Edge for dynamic, adaptive edge maintenance and management. Finally, the authors discuss the challenges regarding Edge intelligence and Intelligent Edge. The main requirement is to design a complete system framework covering data acquisition, service deployment, and the placement of \ac{AI/ML} models %partition of the neural network 
considering processing and network resources.

Pham et al. \cite{Pham-2020} introduced an article presenting the research on the integration of \ac{MEC} with \ac{5G} and beyond. The authors discuss \ac{MEC} for \ac{UAV} communication and the integration between \ac{5GC} and vRAN. The authors explain how \acp{UAV} can improve wireless communications, providing  cost-effective, fast, flexible, and efficient deployments. Moreover, \acp{UAV} can provide on-the-fly communications and establish \ac{LoS} communication links to users in a complementary network for \ac{SAR} emergencies and disaster reliefs. In this context, Pham et al. \cite{Pham-2020} describe two typical scenarios integrating \ac{UAV} and \ac{MEC}. An application considers \acp{UAV} operating as aerial users of the cellular-connected network. In this case, the \ac{MEC} server-based \ac{BS} can provide seamless and reliable wireless communications for \acp{UAV} to improve computation performance. Another application refers to \acp{UAV} working as aerial \acp{BS} and equipped with a \ac{MEC} server. Therefore, \ac{MEC}-enabled \ac{UAV} servers give opportunities for mobile users to offload heavy computation tasks. Finally, the authors highlight several challenges integrating \ac{UAV} with the \ac{MEC} system, such as mobility control and trajectory optimization, communication and computation resource optimization, energy-aware resource allocation, and user grouping and \ac{UAV} association.

Specifically, about \ac{SAR} as an \ac{MC} task, Queralta et al. \cite{AutoSOS} show the research efforts within the AutoSOS project.
%, supported by the Academy of Finland's. 
This project designs an autonomous multi-robot \ac{SAR} assistance platform using \ac{AI} models for object detection. The platform operates in the reconnaissance missions over the sea by executing adaptive \ac{DL} algorithms 
%and  collecting sensors and computational resources 
in \acp{UAV} and boats. Moreover, \acp{UAV} can autonomously reconfigure their spatial arrangement to allow multi-hop communication, for example, when a direct connection between a \ac{UAV} transmitting information and the vessel is unavailable. This reconfiguration uses algorithms for autonomous decision-making at the edge devices, i.e., \acp{UAV}, considering independent task migration and communication priority decisions within the multi-UAV system. In this context, the authors discuss the need to integrate a single multi-agent control loop using \ac{DL} techniques for advanced vision, communication constraints, spatial awareness, and computation distribution. Moreover, Queralta et al. highlight that this integration cannot require high computational resources since a \ac{UAV}'s low energy consumption is needed to increase its autonomy.
%more operation time.

%\begin{table*}[!ht]
% \begin{table*}[htb]
% \centering
% \caption{Related works involving \acp{UAV} assisted by \ac{AI/ML} and focused in \ac{MC} services over \ac{5G}/B5G infrastructures}
% \label{tab:comparison}
% \def\arraystretch{1.2}
% \begin{tabular}{l|c|c|c|c|c|c|c|c}
% \hline
% \textbf{Article} & 
% \textbf{Multi-UAV} & 
% \textbf{Edge} & 
% \textbf{AI} & 
% \textbf{5GC} & 
% \textbf{MEC}  &
% \textbf{vRAN} & 
% \textbf{Cloud} & 
% \textbf{MC Services} \\ \hline

% %%%%%%%%%%%%%%%%%%%%%%%%
% \rowcolor[HTML]{C5DEED} Xu et al (2020) \cite{Dianlei-2020} & 
%  &
% \checkmark & 
% \checkmark & 
%  &
%  &   
%  & 
% \checkmark & 
% \\ \hline

% %%%%%%%%%%%%%%%%%%%%%%%%
% Wang et al (2020) \cite{Wang-2020} & 
%  &
% \checkmark & 
% \checkmark & 
%  &
% \checkmark &   
%  & 
% \checkmark & 
% \\ \hline

% %%%%%%%%%%%%%%%%%%%%%%%%
% \rowcolor[HTML]{C5DEED}Pham et al (2020) \cite{Pham-2020} & 
% \checkmark &
% \checkmark & 
% \checkmark & 
% \checkmark &
% \checkmark &   
% \checkmark & 
% \checkmark & 
% \checkmark \\ \hline

% %%%%%%%%%%%%%%%%%%%%%%%%
% Queralta et at (2020) \cite{AutoSOS} & 
% \checkmark &
% \checkmark & 
% \checkmark & 
%  &
%  &   
%  & 
%  & 
% \checkmark \\ \hline

% %%%%%%%%%%%%%%%%%%%%%%%%
% \rowcolor[HTML]{C5DEED} Zhang et at (2020) \cite{Beyond-D2D} & 
% \checkmark &
%  & 
% \checkmark & 
%  &
%  \checkmark &   
%  \checkmark & 
%  & 
%  \\ \hline

% \end{tabular}
% \end{table*}

\begin{table}[!h]
\centering
\caption{Related work involving \ac{UAV}s assisted by \ac{AI/ML} and focused on \ac{MC} services over 5G/\ac{B5G} infrastructures}
\label{tab:comparison}
\def\arraystretch{1.2}
\begin{tabular}{c|c|c|c|c|c|c|c|c}
\hline

 & 
\textbf{Multi} & 
 & 
 & 
 & 
  &
 & 
 & 
\textbf{MC} \\ 

\multirow{-2}{*}{\textbf{Article}} & 
\textbf{UAV} & 
\multirow{-2}{*}{\textbf{Edge}} & 
\multirow{-2}{*}{\textbf{AI}} &  
\multirow{-2}{*}{\textbf{5GC}} &  
\multirow{-2}{*}{\textbf{MEC}} &  
\multirow{-2}{*}{\textbf{vRAN}} &  
\multirow{-2}{*}{\textbf{Cloud}} &  
\textbf{Services} \\ \hline

%%%%%%%%%%%%%%%%%%%%%%%%
%\rowcolor[HTML]{C5DEED} Xu et al. (2020) \cite{Dianlei-2020} & 
\rowcolor[HTML]{C5DEED} \cite{Dianlei-2020} & 
 &
\checkmark & 
\checkmark & 
 &
 &   
 & 
\checkmark & 
\\ \hline

%%%%%%%%%%%%%%%%%%%%%%%%
%Wang et al. (2020) \cite{Wang-2020} & 
\cite{Wang-2020} & 
 &
\checkmark & 
\checkmark & 
 &
\checkmark &   
 & 
\checkmark & 
\\ \hline

%%%%%%%%%%%%%%%%%%%%%%%%
%\rowcolor[HTML]{C5DEED}Pham et al. (2020) \cite{Pham-2020} & 
\rowcolor[HTML]{C5DEED} \cite{Pham-2020} & 
\checkmark &
\checkmark & 
\checkmark & 
\checkmark &
\checkmark &   
\checkmark & 
\checkmark & 
\checkmark \\ \hline

%%%%%%%%%%%%%%%%%%%%%%%%
%Queralta et al. (2020) \cite{AutoSOS} & 
\cite{AutoSOS} & 
\checkmark &
\checkmark & 
\checkmark & 
 &
 &   
 & 
 & 
\checkmark \\ \hline

%%%%%%%%%%%%%%%%%%%%%%%%
%\rowcolor[HTML]{C5DEED} Zhang et al. (2020) \cite{Beyond-D2D} & 
\rowcolor[HTML]{C5DEED} \cite{Beyond-D2D} & 
\checkmark &
 & 
\checkmark & 
 &
 \checkmark &   
 \checkmark & 
 & 
 \\ \hline

\end{tabular}
\end{table}

Glimpsing a \ac{6G} network, Zhang et al. \cite{Beyond-D2D} propose a \ac{U2X} communication framework. The authors argue that \acp{UAV} are unsuited for achieving a high data rate by directly connecting the terrestrial cellular networks. Therefore, the authors apply three techniques for \ac{U2X} communications: cooperative sensing and transmission protocol, \ac{UAV} trajectory design, and radio resource management considering vRANs. Together, they provide a feasible architecture for \ac{UAV} sensing utilization in the \ac{6G} network. Moreover, the authors discuss the open problems of \ac{U2X} communications, such as \ac{UAV} Cooperation with \ac{U2X} Communications to reduce the cost of power and spectrum resources, as well as \ac{MEC} with \ac{U2X} Communications to minimize the computation workload of the base station and to improve \ac{QoS}.

We summarize this short sample of the state-of-the-art in Table~\ref{tab:comparison}, which indicates the fundamental elements for supporting \ac{UAV}-based critical missions considered in each article, such as  Multi-UAV, edge, \ac{AI}, \ac{MEC}, vRAN, cloud, and \ac{MC} services. According to our literature review, the manuscripts do not discuss the integration of all elements that concern the proposed \ac{SI}. However, the literature clearly indicates the trend towards an intelligent global system with a cognitive control loop using \ac{AI/ML} techniques.

%\textcolor{blue}{Ajustar de acordo com a Section 3...}
\section{Conclusion and Future Work}\label{sec:conclusion}

Standardization and wide adoption of \ac{AI/ML} are key strategies to reach scale, reduce cost, and achieve \ac{MC} services efficiency that heavily depends on communication and computing systems. \acp{UAV} significantly contribute to improving \ac{MC} services but demand standards and efficient \ac{AI/ML} for optimized performance. In this article, we presented \ac{SI}, an architecture for providing full \ac{AI/ML} capabilities for standardized systems supporting \ac{MC} services. In addition to describing the whole architecture and its interaction with the assisted system, we also presented experiments that illustrate \ac{SI} benefits. The initial results show the challenges imposed in some search and rescue scenarios while also offering the potential gains obtained with the introduction of \ac{SI}.

The described experiments are timely for researching 6G and future generation computer systems because the current state-of-the-art in UAV-based has not yet found the requirements for MC tasks, such as the timeliness of data processing. The present research seeks solutions to how UAV can improve their sensing range and how UAVs can become more intelligent through cooperation. Aspects such as delays need to be taken into account, and simulators are rather limited concerning mimicking all variability found in critical missions. 

While the experiments of this article exercised some of the essential components of \ac{SI} architecture, implementing this software that covers all its components is still lacking. Therefore, as future work, we intend to develop and publicly available a functional implementation of \ac{SI} that minimally illustrates all components. Given the size and complexity of this task, we plan to perform it in phases. Each one is complemented by additional use cases that illustrate the components involved and the benefits obtained. Additionally, we are interested in investigating how to evolve the \ac{AI/ML} awareness level of assisted systems. For example, the necessary changes to turn \ac{MEC} at least \ac{AI/ML} basic-aware and turn \ac{5G} core \ac{AI/ML} advanced-aware. In this context, the concept of ML pipeline introduced by the specification ITU-T Y.3172 seems a promising approach for adding awareness to assisted systems.

\section*{Acknowledgment}

This article was conducted with partial financial support from the National Council for Scientific and Technological Development (CNPq) under grant number 130555/2019-3.

%and from the Coordination for the Improvement of Higher Education Personnel (CAPES) - Finance Code 001, Brazil.


\begin{acronym}
\acro{5G}{5th Generation}
\acro{6G}{6th Generation}
\acro{ENI}{Experiential Networked Intelligence}
\acro{AI/ML}{Artificial Intelligence/Machine Learning}
\acro{MC}{Mission Critical}
\acro{B5G}{Beyond 5G}
\acro{AIMET}{AI Model Efficiency Toolkit}
\acro{C2}{Command and Control}
\acro{RAN}{Radio Access Network}
\acro{non-RT}{non-Real-Time}
\acro{near-RT}{near-Real-Time}
%\acro{vRAN}{Virtualized Radio Access Network}
\acro{vRAN}{virtual RAN} %or above? Virtualized?
\acro{DRL}{Deep Reinforcement Learning}
\acro{SI}{System Intelligence}
\acro{ML}{Machine Learning}
\acro{PNI-NPN}{Public Network Integrated NPN}
\acro{SNPN}{Standalone NPN}
\acro{PLMN}{Public Land Mobile Network}
\acro{AN}{Access Network}
\acro{CN}{Core Network}
\acro{NWDAF}{Network Data Analytics Function}
\acro{SON}{Self-Organizing Network}
\acro{NS}{Network Slicing}
\acro{NPN}{Non-Public Network}
\acro{NRF}{Network Repository Function}
\acro{B5G}{Beyond 5G}
\acro{Naf}{Network application function}
\acro{AS}{Assisted System}
\acro{UPF}{User Plane Function}
\acro{3GPP}{3rd Generation Partnership Project}
\acro{5GC}{5G Core}
\acro{ETSI}{European Telecommunications Standards Institute}
\acro{gNB}{Next generation NodeB}
\acro{SBA}{Service-Based Architecture}
\acro{NG-RAN}{Next-Generation Radio Access Network}
\acro{SBI}{Service-Based Interface}
\acro{DNN}{Deep Neural Network}
\acro{DL}{Deep Learning}
\acro{FPS}{Frames per Second}
\acro{UAV}{Unmanned Aerial Vehicle}
%\acro{UAVs}{Unmanned Aerial Vehicles}
\acro{EI}{Edge Intelligence}
\acro{eMBB}{enhanced Mobile Broadband }
\acro{URLL}{ultra-Reliable Low Latency}
\acro{SAR}{Search and Rescue}
\acro{SDAR}{Search, Diagnostic, and Rescue}
\acro{BS}{Base Station}
\acro{AI}{Artificial Intelligence}
\acro{MEC}{Multi-access Edge Computing}
\acro{MIMO}{Multiple Input Multiple Output}
\acro{NF}{Network Function}
\acro{NFV}{Network Functions Virtualization}
\acro{SDN}{Software-Defined Networking}
\acro{SDR}{Software-Defined Radio}
\acro{VNF}{Virtual Network Function}
\acro{USRP}{Universal Software Radio Peripherals}
\acro{VM}{Virtual Machine}
\acro{SRS}{Software Radio Systems}
\acro{UE}{User Equipment}
\acro{BBU}{Base Band Unit}
\acro{RRH}{Remote Radio Head}
\acro{QoS}{Quality of Service}
\acro{VNF}{Virtualized Network Functions}
\acro{RIC}{RAN Intelligent Controller}
\acro{C-RAN}{Cloud Radio Access Network}
\acro{O-RAN}{Open Radio Access Network}
\acro{O-RU}{O-RAN Radio Unit}
\acro{CU}{Central Unit}
\acro{O-DU}{O-RAN Distributed Unit}
\acro{LTE}{Long Term Evolution}
\acro{EPC}{Evolved Packet Core}
\acro{gNB}{5G base-station}
\acro{MME}{Mobility Management Entity}
\acro{S-/P-GW}{Serving-/Packet Data Network-Gateway}
\acro{HSS}{Home Subscriber Server}
\acro{FFT}{Fast Fourier Transform}
\acro{MAC}{Media Access Control}
\acro{API}{Application Programming Interface}
\acro{IoT}{Internet of Things}
\acro{TDD}{Time Division Duplex}
\acro{OAI}{OpenAirInterface}
\acro{OS}{Operating System}
\acro{PHY}{Physical Layer}
\acro{CR}{Cognitive Radio}
\acro{IAB}{Integrated Access Backhaul}
\acro{CAPEX}{CAPital EXpenditures}
\acro{OPEX}{OPerational EXpenditures}
\acro{ITU-T}{International Telecommunication Union}
\acro{APN}{Access Point Name}
\acro{CS}{Control Signal}
\acro{LoS}{Line-of-Sight}
\acro{U2X}{UAV-to-Everything}
\end{acronym}

\bibliographystyle{elsarticle-num-names}
\bibliography{ref.bib}

\end{document}